\documentclass[9pt,twocolumn,twoside]{pnas-new}
\templatetype{pnasresearcharticle} % Choose template 
\title{Discovery in space of ethanolamine, the simplest phospholipid head group
}

\author[a,b,1]{Víctor M. Rivilla}
\author[a]{Izaskun Jiménez-Serra} 
\author[a]{Jesús Martín-Pintado}
\author[a]{Carlos Briones}
\author[a]{Lucas F. Rodríguez-Almeida}
\author[a]{Fernando Rico-Villas}
\author[c]{Belén Tercero}
\author[d]{Shaoshan Zeng}
\author[a,b]{Laura Colzi}
\author[c]{Pablo de Vicente}
\author[e,f]{Sergio Martín}
\author[g,h]{Miguel A. Requena-Torres}

\affil[a]{Centro de Astrobiología (CSIC-INTA), Ctra. de Ajalvir Km. 4, 28850, Torrejón de Ardoz, Madrid, Spain}
\affil[b]{INAF-Osservatorio Astrofisico di Arcetri, Largo Enrico Fermi 5, 50125, Florence, Italy}
\affil[c]{Observatorio Astronómico Nacional (OAN-IGN), Calle Alfonso XII, 3, 28014 Madrid, Spain}
\affil[d]{Star and Planet Formation Laboratory, Cluster for Pioneering Research, RIKEN, 2-1 Hirosawa, Wako, Saitama, 351-0198, Japan}
\affil[e]{European Southern Observatory, ALMA Department of Science, Alonso de Córdova 3107, Vitacura 763 0355, Santiago, Chile}
\affil[f]{Joint ALMA Observatory, Department of Science Operations, Alonso de Córdova 3107, Vitacura 763 0355, Santiago, Chile}
\affil[g]{University of Maryland, Department of Astronomy, College Park, ND 20742-2421, USA}
\affil[g]{Department of Physics, Astronomy and Geosciences, Towson University, Towson, MD 21252, USA}

\leadauthor{Rivilla} 

\significancestatement{The detection of ethanolamine (NH$_2$CH$_2$CH$_2$OH) in a molecular cloud in the interstellar medium confirms that a precursor of phospholipids is efficiently formed by interstellar chemistry. Hence, ethanolamine could have been transferred from the proto-Solar nebula to planetesimals and minor bodies of the Solar System, and thereafter to our planet. The prebiotic availability of ethanolamine on early Earth could have triggered the formation of efficient and permeable amphiphilic molecules such as phospholipids, playing thus a relevant role in the evolution of the first cellular membranes needed for the emergence of life.
}

\authorcontributions{Author contributions: V.M.R., J.M.P., and F.R. conducted the IRAM 30m observations. B.T. and P. dV. performed the Yebes 40m observations. S.M., J.M.P. , V.M.R., and F.R. contributed to the development of the software used in the analysis. V.M.R., J.M.P., L.F. R-A., and I.J-S reduced the data of the full observational survey. V.M.R. identified ethanolamine and related species in the observed spectra, and performed the analysis presented in this work. L.F. R-A., L.C, S.Z., J.M.P., I.J-S. significantly contributed to the identification of other molecular species in the spectral survey. C.B. investigated the biochemical relevance of this detection. V.M.R. wrote the initial draft, and prepared the figures; I.J-S., J.M.P. and C.B. revised the first version of the manuscript and contributed significantly to write the final version. All authors reviewed the manuscript, contributed actively to the scientific discussion, and to finalizing the manuscript. V.M.R. and I.J-S. acquired the financial support for the scientific projects that led to this publication. 
}
\authordeclaration{The authors declare no competing interests.}

\correspondingauthor{\textsuperscript{1}To whom correspondence should be addressed. E-mail: vrivilla@cab.inta-csic.es}


\begin{abstract}
Cell membranes are a key element of life because they keep the genetic material and metabolic machinery together. All present cell membranes are made of phospholipids, yet the nature of the first membranes and the origin of phospholipids are still under debate. We report here the first detection in space of ethanolamine, NH$_2$CH$_2$CH$_2$OH, which forms the hydrophilic head of the simplest and second most abundant phospholipid in membranes. The molecular column density of ethanolamine in interstellar space is $N$=(1.51$\pm$0.07)$\times$10$^{13}$ cm$^{-2}$, implying a molecular abundance with respect to H$_2$ of (0.9-1.4)$\times$10$^{-10}$. Previous studies reported its presence in meteoritic material but they suggested that it is synthesized in the meteorite itself by decomposition of amino acids.  However, we find that the proportion of the molecule with respect to water in the interstellar medium is similar to the one found in the meteorite (10$^{-6}$).
These results indicate that ethanolamine forms efficiently in space and, if delivered onto early Earth, it could have contributed to the assembling and early evolution of primitive membranes.

\end{abstract}

\dates{This manuscript was compiled on \today}
\doi{\url{www.pnas.org/cgi/doi/10.1073/pnas.XXXXXXXXXX}}

\begin{document}

\maketitle
\thispagestyle{firststyle}
\ifthenelse{\boolean{shortarticle}}{\ifthenelse{\boolean{singlecolumn}}{\abscontentformatted}{\abscontent}}{}

\begin{figure*}%[tbhp]
\centering
\includegraphics[width=0.85\linewidth]{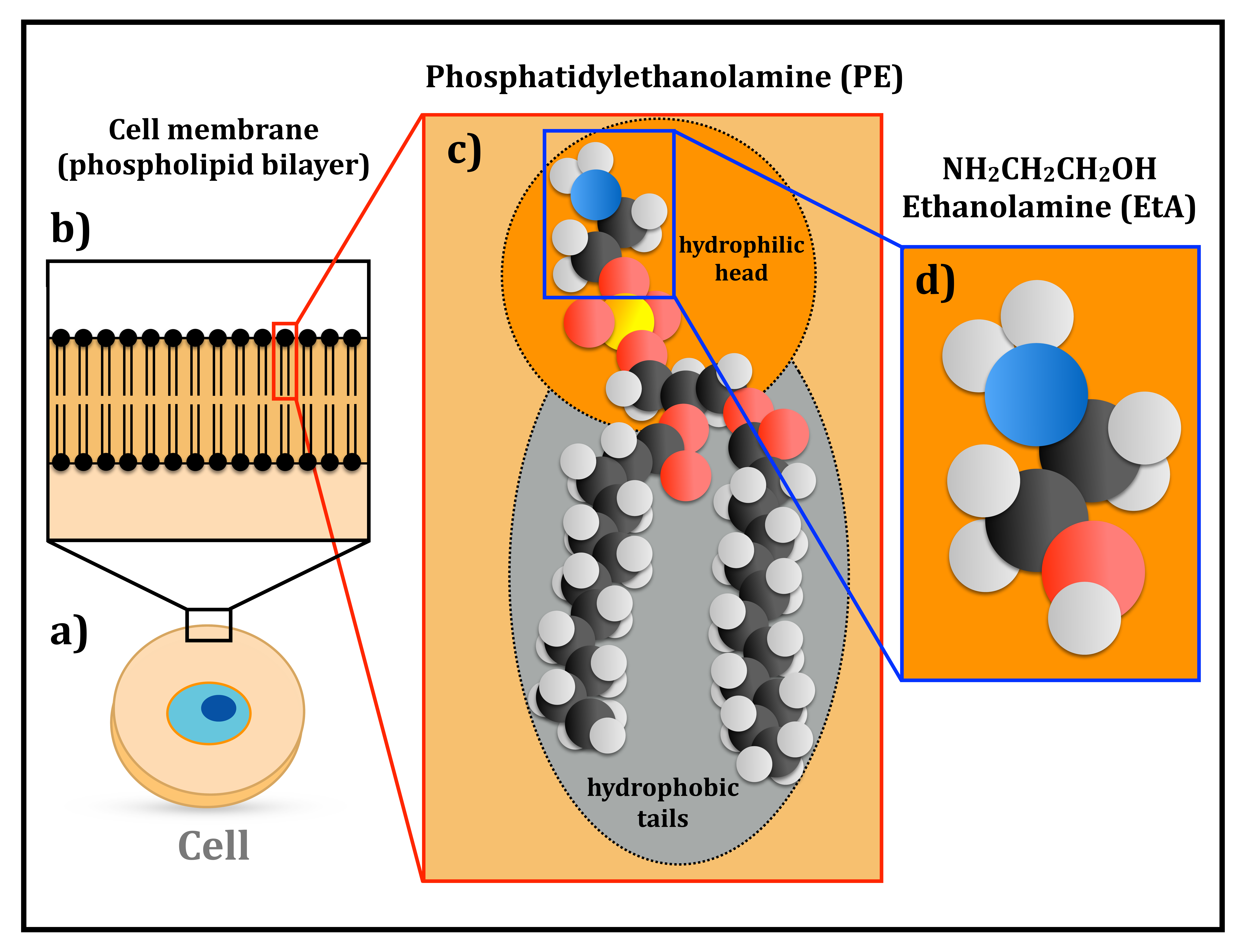}
\caption{Structure of cellular membranes.  a) Schematic view of a cell; b) Zoom-in view of the cell membrane, formed by a phospholipid bilayer; c) Three-dimensional structure of the phospholipid phosphatidylethanolamine (PE), formed by a hydrophilic head composed of ethanolamine, a phosphate group linked to glycerol, and two hydrophobic fatty-acid tails (black, red, blue and white balls denote carbon, oxygen, nitrogen and hydrogen atoms, respectively); d) Ethanolamine (EtA), the molecular species detected in space for the first time and reported in this work.}
\label{fig:membranes}
\end{figure*}

\dropcap{L}ife is based on three basic subsystems: a compartment, a metabolic machinery, and information-bearing molecules together with replication mechanisms \cite{szostak2011,delaescosura2015}. Among these key elements, compartmentalization is a fundamental prerequisite in the process of the emergence and early evolution of life \cite{deamer2005,sole2009}. Indeed, cellular membranes encapsulate and protect the genetic material, as well as enable the metabolic activities within the cell.  The membranes of all current cells are made of a bilayer of phospholipids (Fig. \ref{fig:membranes}a$\&$b), which are composed of a polar hydrophilic head (an alcohol phosphate group combined with a head group such as ethanolamine, choline or serine), and two non-polar hydrophobic tails (hydrocarbon chains derived from fatty acids), as depicted in Fig. \ref{fig:membranes}c. 

The process through which the first phospholipids were formed remains unknown. Initial works proposed that phospholipids could be synthesized under possible prebiotic conditions \cite{hargreaves1977,oro1978,rao1987}, but the availability of the precursor molecules on early Earth was questioned \cite{deamer2005,ruiz-mirazo2017}. Alternatively, the building blocks of phospholipids could have been delivered from space. A broad repertoire of prebiotic molecules could have been provided to the early Earth through the bombardment of comets and meteorites \cite{chyba1992,pizzarello2010}. Laboratory impact experiments \cite{bertrand2009,mccaffrey2014} have demonstrated that a significant fraction of the prebiotic molecules in comets and meteorites can survive both passage through the planetary atmosphere and the impact on the surface.

In particular, some structural parts of phospholipids are known to be present in meteorites, such as fatty acids, alcohols and phosphonic acids \cite{pizzarello2010,septhon2002,cooper1992}. The glycerol phosphate group has been shown to be synthesized in irradiation experiments of interstellar ice analogues \cite{layssac2020,zhu2020}, which supports the idea that they can form in space. Regarding the different head groups of phospholipids, ethanolamine (also known as glycinol or 2-aminoethanol, NH$_2$CH$_2$CH$_2$OH, Fig. \ref{fig:membranes}d), is the simplest one, and it forms the second most abundant phospholipid in biological membranes: phosphatidylethanolamine (PE, see Fig. \ref{fig:membranes}c). In addition, ethanolamine has been proposed as a direct precursor of the simplest amino acid, glycine (NH$_2$CH$_2$COOH), in simulated archean alkaline hydrothermal vents \cite{zhang2017}, considered as one of the likely environments for the origin of life \cite{sojo2016}.

Ethanolamine (hereafter EtA) has been found in the Almahata Sitta meteorite \cite{glavin2010}, yet its origin is not known. A possible chemical formation route was proposed to be the thermal decomposition of amino acids under specific unusual conditions in the parent asteroid. This would limit the availability of EtA in the early Earth for the formation of phospholipids and thereafter of cell membranes. Another possibility is that EtA is formed from smaller interstellar precursors. However, the detection of EtA in the interstellar medium (ISM) has remained so far elusive \cite{wirstrom2007}.

\subsection*{Results}

\begin{figure*}%[tbhp]
\centering
\includegraphics[scale=0.55]{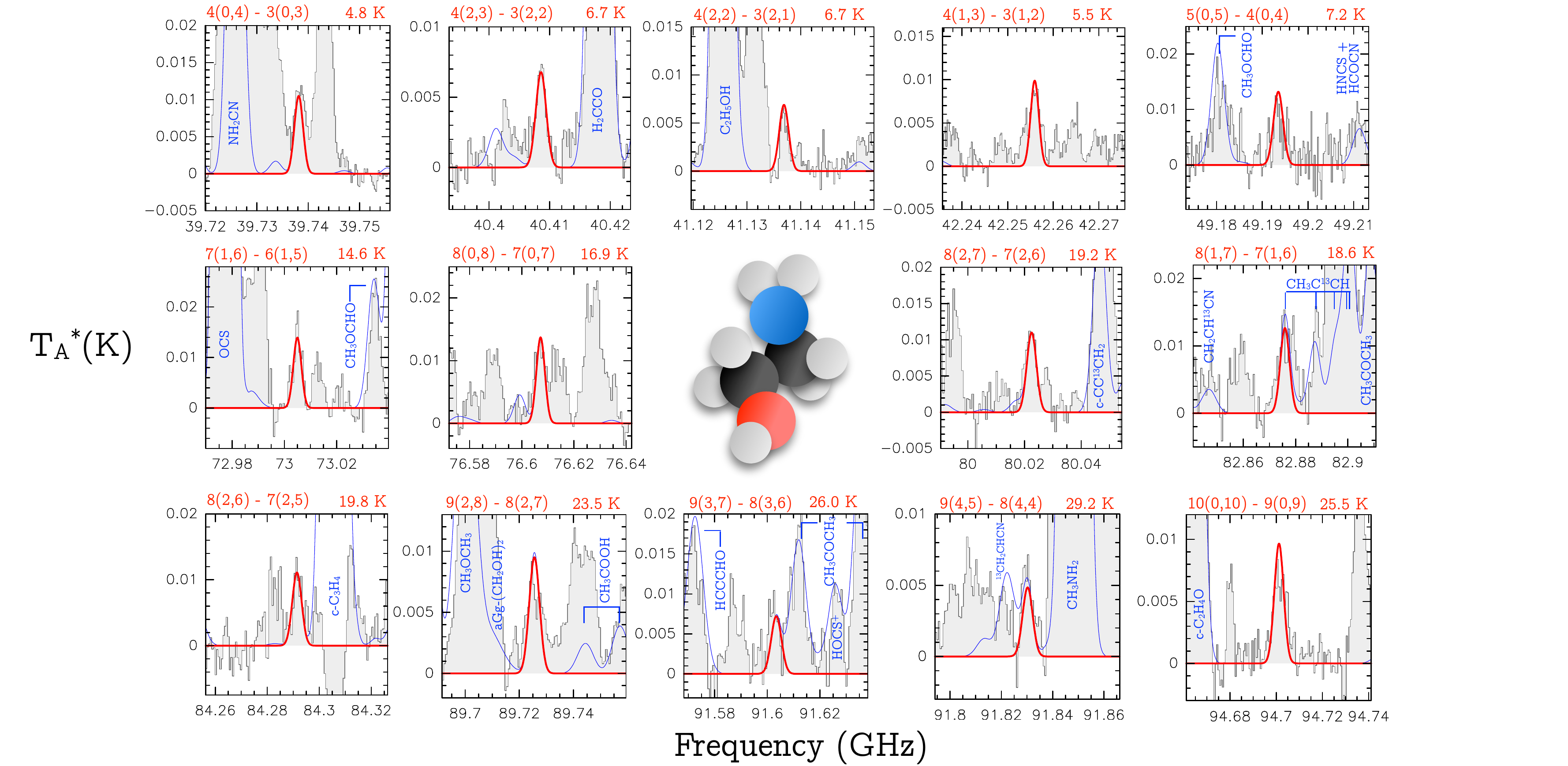}
\caption{Unblended or slightly blended transitions of EtA towards the G+0.693-0.027 molecular cloud. The quantum numbers involved in the transition are indicated in the upper left of each panel, and the energies of the upper level are indicated in the upper right. The red thick line depicts the best LTE fit to the EtA rotational transitions. The thin blue line shows the expected molecular emission from all the molecular species identified in our spectral survey, overplotted to the observed spectra (gray histograms). The three-dimensional structure of EtA is shown in the center of the figure: black, red, blue and white balls denote carbon, oxygen, nitrogen and hydrogen atoms, respectively. }
\label{fig:eta-detections}
\end{figure*}

We have detected EtA towards the molecular cloud G+0.693-0.027 (hereafter G+0.693), located in the SgrB2 complex in the Galactic Center, as shown in Fig. \ref{fig:eta-detections}. This region is one of the most chemically rich reservoirs of molecules in the Galaxy, with a plethora of organic species detected \cite{requena-torres2008,zeng2018,rivilla2019,rivilla2020,jimenez-serra2020}. The extremely rich gas-phase chemical composition of this region is due to erosion of the ice mantles of interstellar dust grains by large-scale low-velocity ($<$ 20 km s$^{-1}$) shocks \cite{martin2008} induced by a collision between massive molecular clouds \cite{zeng2020}.
For the typical (intermediate) densities of G+0.693 of a few 10$^{4}$ cm$^{-3}$ \cite{zeng2020}, the emission is sub-thermally excited, yielding very low excitation temperatures ($T_{\rm ex}$) in the range 5$-$15 K \cite{requena-torres2008,zeng2018}. Since only low-energy molecular transitions are excited, the density of molecular lines is substantially lower than in hotter sources such as massive molecular hot cores or low-mass hot corinos, alleviating the problems of line blending and line confusion. This, along with the effect of shock-induced desorption of interstellar ices, makes G+0.693 an excellent target for the detection of new molecular species in the ISM.

We analyzed the molecular data of a high sensitivity unbiased spectral survey carried out with the IRAM 30m and the Yebes 40m radiotelescopes. Detailed information about the observations is presented in the Materials and Methods section. The identification of the rotational transitions of EtA was performed using the SLIM (Spectral Line Identification and Modeling) tool within the MADCUBA package \cite{martin2019}. We predicted the synthetic spectrum of EtA under the assumption of Local Thermodynamic Equilibrium (LTE) conditions. Among the numerous (23,655) transitions of EtA that fall in the spectral range covered by the survey, only tens of them are expected to be excited considering the low excitation temperatures measured in G+0.693 ($T_{\rm ex}\sim$5$-$15 K) \cite{requena-torres2008,zeng2018}. 

We have detected the 45 brightest transitions of EtA, as predicted by the LTE simulation (with line intensities $T_{\rm A}^*$ $>$ 5 mK), 14 out of which appear either unblended or slightly blended with emission from other molecules. These transitions are shown in Fig. \ref{fig:eta-detections}, and their spectroscopic information is provided in Table \ref{tab:unblended}. The remaining 31 transitions are consistent with the observed spectra but appear blended with brighter emission lines from other molecular species already identified in this molecular cloud (see below). These transitions are shown in Fig. \ref{fig:blended}, and listed in Table \ref{tab:blended}. 

To confirm that the spectral lines detected at the frequencies of the transitions of EtA are not produced by any other molecule, we have performed an extensive search for molecular species in our spectral survey, which includes all the species detected so far in the ISM \cite{mcguire2018}, and all other species reported towards G+0.693 in previous works \cite{requena-torres2008,zeng2018,rivilla2019,rivilla2020,jimenez-serra2020}. The predicted contribution from all molecular species is shown with a blue solid line in Fig. \ref{fig:eta-detections}, confirming that 14 transitions of EtA are either clean or not significantly contaminated by the emission from other molecules. We have used these 14 transitions to perform the LTE fit and to derive the physical parameters of the emission of EtA. 
We used the AUTOFIT tool of MADCUBA$-$SLIM, which finds the best agreement between the observed spectra and the predicted LTE model (see details in the Materials and Methods section). To perform the fit we have considered not only the emission of EtA, but also the predicted emission from all the species identified in the region (blue line in Fig. \ref{fig:eta-detections}). The best fitting LTE model for EtA gives a molecular column density of $N$=(1.51$\pm$0.07)$\times$10$^{13}$ cm$^{-2}$, an excitation temperature of $T_{\rm ex}$=10.7$\pm$0.7 K, and a velocity of $v_{\rm LSR}$=68.3$\pm$0.4 km s$^{-1}$ (the linewidth was fixed to 15 km s$^{-1}$, see details in the Materials and Methods section). The derived $T_{\rm ex}$ and $v_{\rm LSR}$ are very similar to those from other species previously analyzed in G+0.693 \cite{requena-torres2008,zeng2018,rivilla2019,rivilla2020,jimenez-serra2020}. To derive the abundance of EtA with respect to molecular hydrogen, we have used the H$_2$ column density inferred from observations of C$^{18}$O \cite{martin2008}, obtaining a value in the range (0.9$-$1.4)$\times$10$^{-10}$.

\begin{figure}%[tbhp]
\centering
\includegraphics[width=.9\linewidth]{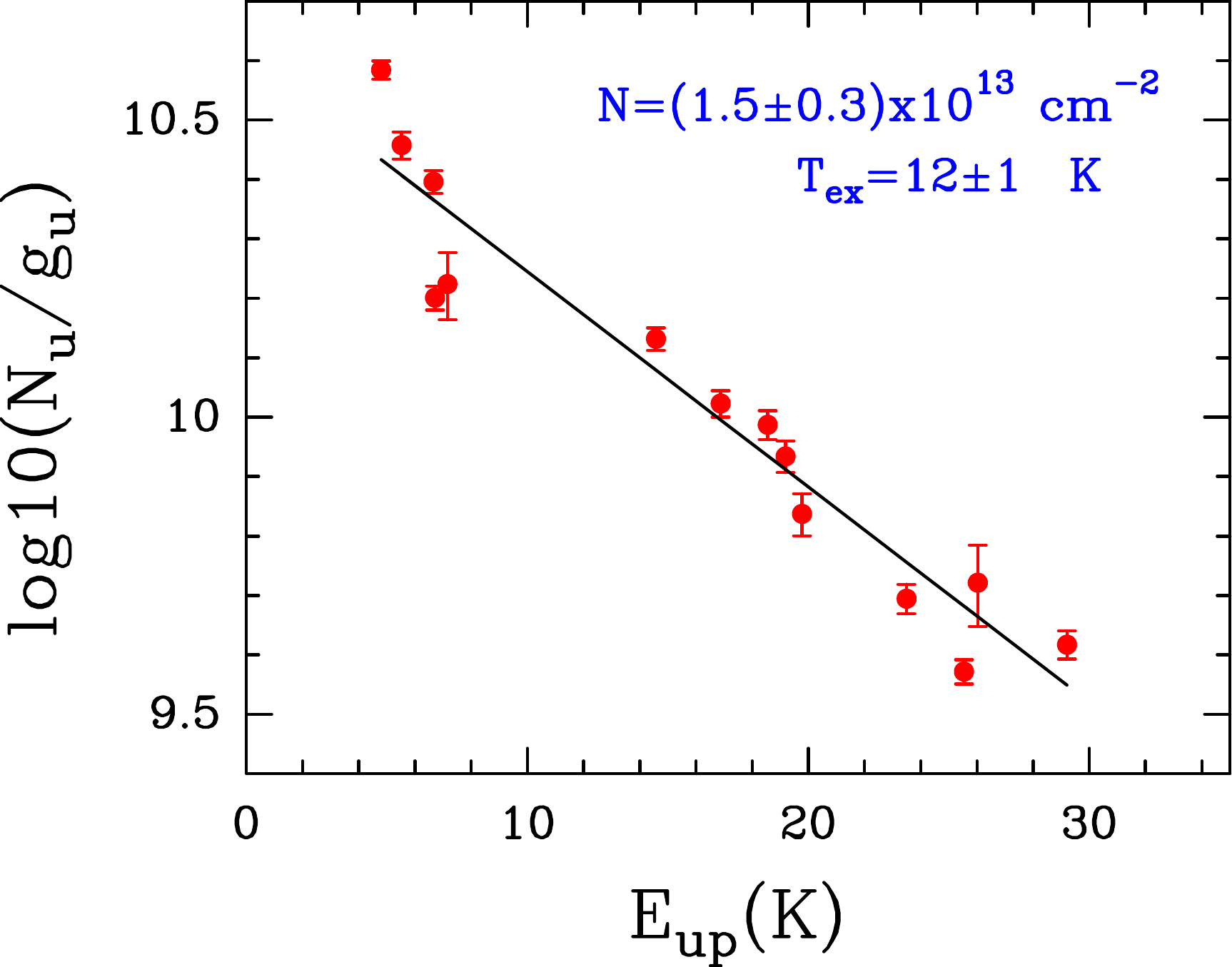}
\caption{Rotational diagram of EtA. The analysis procedure is described in the Materials and Methods section. The red dots correspond to the 14 EtA transitions shown in Fig. \ref{fig:eta-detections} and Table \ref{tab:unblended}. The black line is the best linear fit to the data points. The derived values for the molecular column density ($N$) and the excitation temperature ($T_{\rm ex}$), along with their uncertainties, are indicated in blue in the upper right corner.}
\label{fig:RD}
\end{figure}

We have also performed a complementary analysis using the rotational diagram method implemented in MADCUBA (see further description in the Materials and Methods section). Fig. \ref{fig:RD} shows the rotational diagram obtained using the 14 EtA transitions from Fig. \ref{fig:eta-detections}. We derived physical parameters fully consistent with the MADCUBA$-$AUTOFIT analysis: $N$=(1.5$\pm$0.3)$\times$10$^{13}$ cm$^{-2}$, and $T_{\rm ex}$=12$\pm$1 K.

\begin{figure*}%[tbhp]
\centering
\includegraphics[scale=0.55]{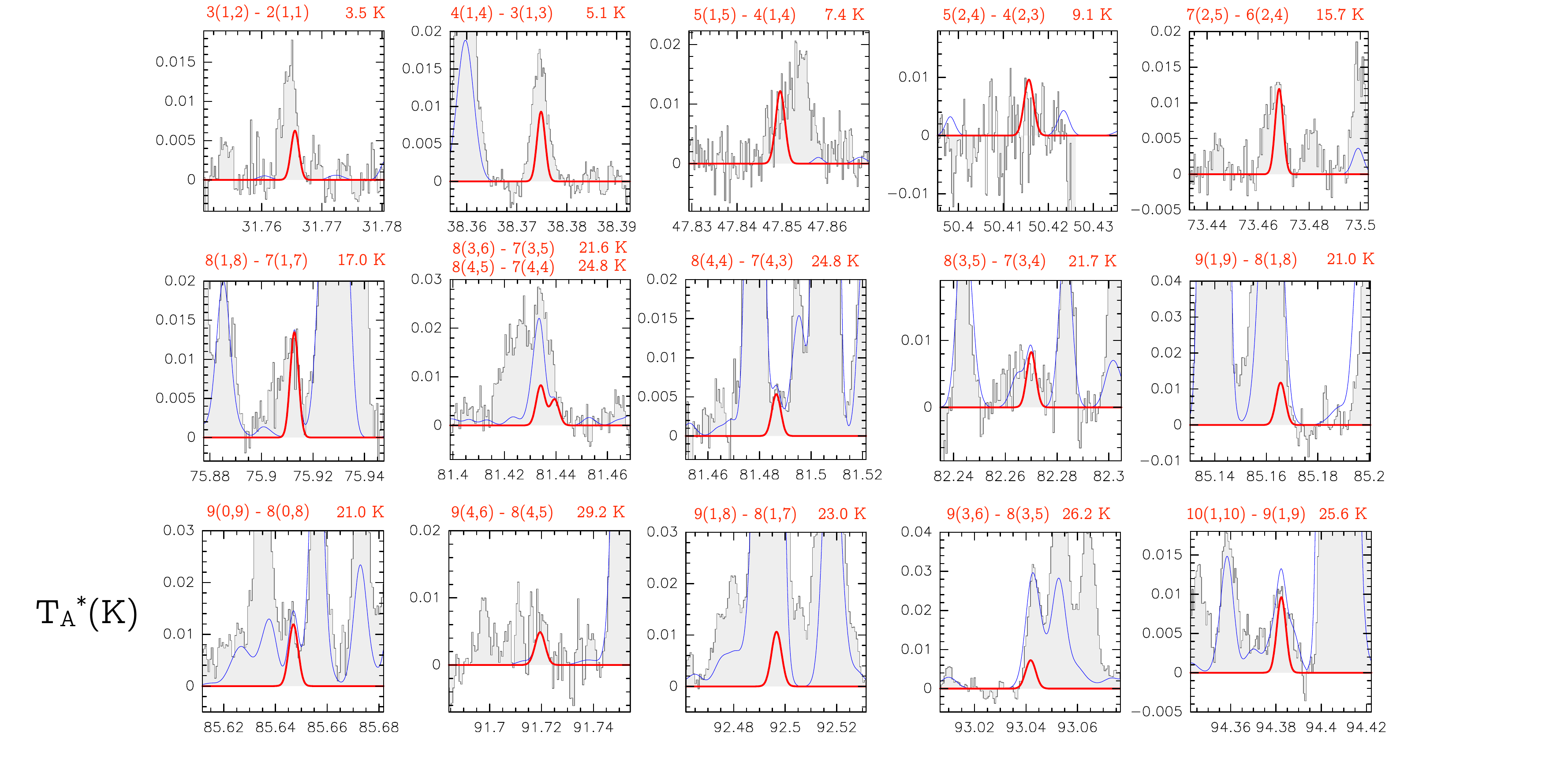}
\includegraphics[scale=0.55]{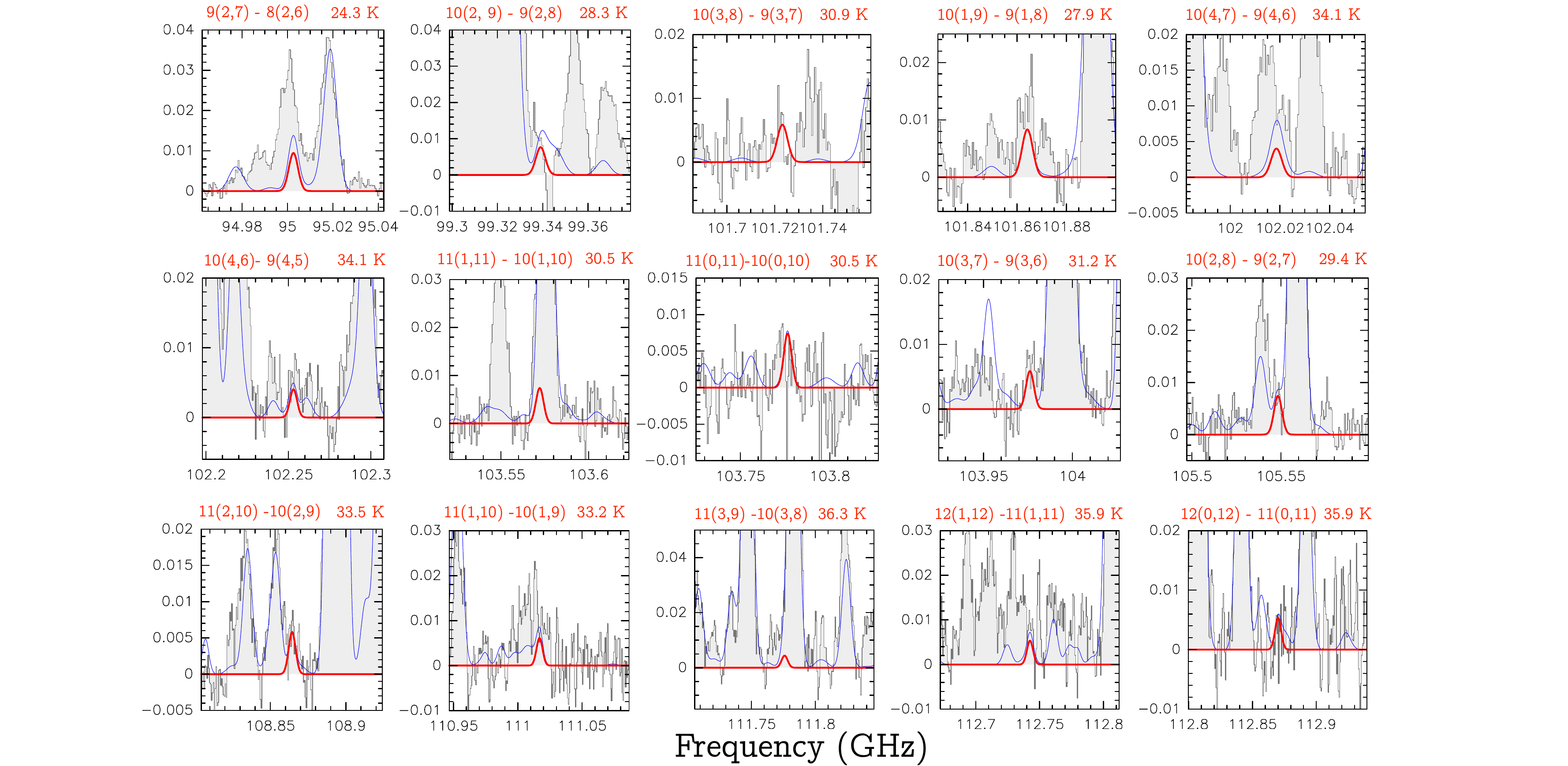}
\caption{Transitions of ethanolamine that appear blended in the observed spectra of G+0.693. All of them have line intensities $>$ 5 mK, according to the best LTE fit described in the text. The quantum numbers involved in the transition are indicated in the upper left of each panel, and the energies of the upper level are indicated in the upper right. The red thick line depicts the best LTE fit obtained fitting the EtA rotational transitions shown in Fig. \ref{fig:eta-detections}. The thin blue line shows the predicted molecular emission from all the molecular species identified in our spectral survey, overplotted to the observed spectra (gray histograms).}
\label{fig:blended}
\end{figure*}

\begin{figure*}%[tbhp]
\centering
\includegraphics[scale=0.26]{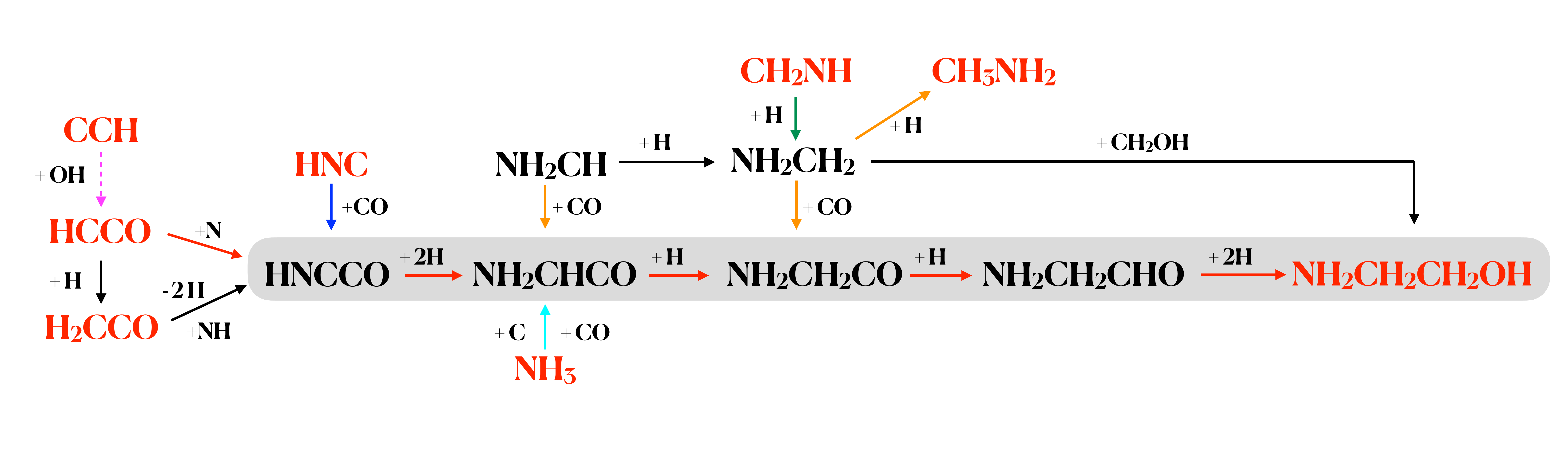}
\vspace{-10mm}
\caption{Summary of the chemical routes proposed for the formation of EtA in the ISM. The molecular species in red have been detected towards the G+0.693 molecular cloud. The gray shaded area corresponds to a hydrogenation chain. The chemical reactions indicated with colored arrows have been proposed in previous works:  magenta \cite{wakelam2015}, blue \cite{kameneva2017}, orange \cite{suzuki2018}, cyan \cite{krasnokutski2020}, green \cite{ruaud2015,suzuki2018}. In black, we show the formation routes proposed in this work. The solid arrows indicate surface chemistry reactions, and dashed arrows denote gas phase chemistry.}
\label{fig:chem-network}
\end{figure*}

\subsection*{Discussion}

We report a clear detection in the ISM of EtA, a precursor of phospholipids, with a relatively high abundance (10$^{-10}$ with respect to molecular hydrogen). This detection adds to that of precursors of ribonucleotides \cite{rivilla2019,rivilla2020,jimenez-serra2020} and amino acids \cite{belloche2008,zeng2019} in the ISM. The building blocks of the three subsystems of life could therefore have been synthesized by interstellar chemistry, being part of the natal material that formed the Solar System.

The formation routes of EtA in the ISM are however poorly known. Grain-surface formation of EtA has been demonstrated by laboratory experiments of ultraviolet (UV) irradiation of interstellar ice analogues \cite{bernstein2002}. In these experiments, photolysis of H$_2$O:CH$_3$OH:NH$_3$:HCN ices with a 20:2:1:1 mixture yields EtA as well as other prebiotic species such as the amino acids glycine, alanine and serine. However, the detailed routes that result into the formation of EtA are still not understood.
We discuss here several possible chemical pathways for the formation of EtA in the ISM, which are summarised in Fig. \ref{fig:chem-network}. 

To our knowledge, the only route proposed in the literature \cite{charnley2001,ehrenfreund2001} is the hydrogenation chain of HNCCO on dust grain surfaces (see gray-shaded area in Fig. \ref{fig:chem-network}). 
HNCCO could be formed on grains by  
N-addition to ketenyl (HCCO) \cite{charnley2001}.
HCCO is rarely found in the ISM, with only two detections reported towards the cold dark clouds Lupus-1A and L483 \cite{agundez2015}. We have also searched for HCCO in G+0.693, and tentatively detected it. The details about this detection are provided in the Materials and Methods section. We obtain a column density of $\sim$0.5$\times$10$^{13}$ cm$^{-2}$, a factor $\sim$3 lower than that of EtA. HCCO is expected to be a highly reactive radical on dust grains. This would result in a low ice abundance of HCCO, and consequently also in the gas phase, due to shock-induced sputtering desorption. Alternatively, the amount of HCCO detected towards G+0.693 might have been produced directly in the gas phase through the reaction CCH + OH $\rightarrow$ HCCO + H proposed by \cite{wakelam2015}, since CCH is highly abundant in this cloud \cite{bizzocchi2020}. 

HNCCO could also be formed on dust grains from ketene (H$_2$CCO), after two hydrogen abstractions, and reaction with the imine radical NH (Fig. \ref{fig:chem-network}). 
G+0.693 presents a variety of imines with relatively high abundances \cite{zeng2018,rivilla2019,bizzocchi2020}, which confirms that imine radicals are available on grain surfaces. This route is plausible since ketene is abundant towards G+0.693 \cite{requena-torres2008} with a column density of $N$=2.9$\times$10$^{14}$ cm$^{-2}$, a factor of $\sim$20 larger than that of EtA. 
Alternatively, the formation of HNCCO on grains could proceed as proposed by \cite{kameneva2017} (Fig. \ref{fig:chem-network}) through the combination of HNC and CO, species expected to be abundant on grain surfaces. 

The subsequent hydrogenation of HNCCO can form NH$_2$CHCO (Fig. \ref{fig:chem-network}). This species might also form through other surface-chemistry routes. \cite{krasnokutski2020} proposed a barrierless reaction between NH$_3$, CO and atomic C (Fig. \ref{fig:chem-network}). Given that 3-body reactions are less efficient that 2-body reactions, this route could contribute to the formation of NH$_2$CHCO only if a relatively high abundance of atomic C is available. Since it has been observed that the abundance of C is indeed large in Galactic Center molecular clouds, around half of that of CO \cite{tanaka2011}, and considering that C is expected to be highly reactive, this route might be indeed viable in G+0.693.
We note that the barrierless  NH$_2$CH + CO reaction proposed by \cite{suzuki2018} might also contribute to the formation of NH$_2$CHCO (Fig. \ref{fig:chem-network}).

The hydrogenation of NH$_2$CH could yield the NH$_2$CH$_2$ radical, which might be a key precursor of EtA.
In recent laboratory experiments \cite{ioppolo2020,fedoseev2015} of the non-energetic formation of simple amino acids and sugars under pre-stellar conditions, intermediate radicals such as NH$_2$CH$_2$ and CH$_2$OH are efficiently formed in the hydrogenation reactions towards methylamine (CH$_3$NH$_2$) and methanol (CH$_3$OH). These radicals represent the structural units of EtA and hence, this species could be produced by the non-diffusive reaction between NH$_2$CH$_2$ and CH$_2$OH on the surface of dust grains (Fig. \ref{fig:chem-network}). Similar radical-radical reactions have been proposed as viable routes to form other complex species in the ISM \cite{garrod2006,garrod2008,garrod2013}.
NH$_2$CH$_2$ is expected to be present on the dust grains of G+0.693 since it is an intermediate product between methanimine (CH$_2$NH) and methylamine (CH$_3$NH$_2$) \cite{ruaud2015,suzuki2018,ioppolo2020}, two species that are abundant in G+0.693 \citep{zeng2018}. 

Moreover, NH$_2$CH$_2$ could react with CO, as proposed by \cite{suzuki2018}, to form NH$_2$CH$_2$CO, which can be hydrogenated to form EtA (Fig. \ref{fig:chem-network}). Unfortunately, there is no rotational spectroscopy available for HNCCO, NH$_2$CHCO, or NH$_2$CH$_2$CO so we cannot search for any of these possible precursors of EtA in the G+0.693 spectral survey.

The penultimate step of the hydrogenation chain that results into EtA is aminoacetaldehyde (NH$_2$CH$_2$CHO). The rotational spectra of this species has been studied theoretically by \cite{redondo2020}, although the accuracy of the predicted frequencies ($\sim$0.2$\%$) is still not high enough for any reliable identification in the ISM. Our detection of EtA towards G+0.693 makes NH$_2$CH$_2$CHO a promising species for future interstellar searches, and should motivate new laboratory works to obtain its microwave rotational spectrum with higher accuracy.

The detection of EtA reported in this work with an abundance of $\sim$10$^{-10}$ with respect to H$_2$ enables a rough  comparison with the concentration of this species measured in meteoritic material \cite{glavin2010}. Considering that the abundance of water in the ISM is of the order of 10$^{-4}$ \cite{whittet1991}, the EtA/H$_2$O abundance ratio measured in G+0.693 is of the order of 10$^{-6}$. The Almahata Sitta meteorite, where EtA was detected \cite{glavin2010}, has been classified as a ureilite with a anomalously high fraction of other materials, being the enstatite chondrites (EC) the most abundant \cite{goodrich2015}. Interestingly, EC meteorites have recently been proposed as the origin source of most of Earth’s water \cite{piani2020}. Therefore, meteorites such as Almahata Sitta could have simultaneously delivered to Earth not only water but also prebiotic chemicals such as EtA.  From the concentration of EtA measured in the Almahata Sitta meteorite of 20 ppb \cite{glavin2010}, and the average concentration of water in EC meteorites ($\sim$7500 ppm) \cite{piani2020}, we derive a meteoritic EtA/H$_2$O abundance ratio of 3$\times$10$^{-6}$. This value is consistent with that derived in the ISM. Although isotopic analysis of EtA would be needed to confirm its interstellar origin in meteorites, our results suggest that phospholipid precursors such as EtA formed in the ISM could have been stored in planetesimals and minor bodies of the Solar System, to be subsequently transferred to early Earth.

Once EtA was available on Earth’s surface, it could form phospholipids (in particular PE, see Fig. \ref{fig:membranes}c) under plausible early Earth conditions, as proposed by ref. \cite{oro1978}, and confirmed by prebiotic experiments \cite{rao1987}. It is commonly assumed that the first cell membranes could have been composed of amphiphilic molecules such as fatty acids/alcohols, which are chemically simpler than phospholipids \cite{deamer2005,ruiz-mirazo2017}. However, the availability of EtA in an early Earth could have enabled the progressive replacement of fatty acids/alcohols by more efficient and permeable amphiphilic molecules such as phospholipids. In this scenario, the protocells could have been able to incorporate from the environment the precursor molecules required to start the synthesis of ribonucleic acid (RNA) and eventually other polymeric molecules \cite{monnard2002,budin2011} needed for the first replicative and metabolic processes of life. This has important implications not only for theories of the origin of life on Earth, but also on other habitable planets and satellites anywhere in the Universe.

\begin{table}%[tbhp]
\centering
\caption{Spectroscopic information (rest frequency, Einstein coefficients ($A_{\rm ul}$), and energy of the upper levels ($E_{\rm up}$) of the 14 unblended or slightly blended rotational transitions of EtA detected towards the G+0.693 molecular cloud (shown in Fig. \ref{fig:eta-detections}).}
\begin{tabular}{cccc}
Frequency (GHz) & Transition & log$A_{\rm ul}$ (s$^{-1}$)  & $E_{\rm up}$ (K) \\
\midrule
39.7379429	&  4(0,4) - 3(0,3) &	-5.64603 &	4.8 \\
40.4083769	&  4(2,3) - 3(2,2) &    -5.74549 &	6.7 \\
41.1366268  &  4(2,2) - 3(2,1) &	-5.72208 &	6.7 \\
42.2557255	&  4(1,3) - 3(1,2) &	-5.59088 &	5.5 \\
49.1932727	&  5(0,5) - 4(0,4) &	-5.35979 &	7.2 \\
73.0048603	&  7(1,6) - 6(1,5) &	-4.84093 &	14.6 \\
76.6071016	&  8(0,8) - 7(0,7) &	-4.76916 &	16.9 \\
80.0223886	&  8(2,7) - 7(2,6) &	-4.73500 &	19.2 \\
82.8757878	&  8(1,7) - 7(1,6) &	-4.67140 &	18.6 \\
84.2912932	&  8(2,6) - 7(2,5) &	-4.66443 &	19.8 \\
89.7254251	&  9(2,8) - 8(2,7) &	-4.57750 &	23.5 \\
91.6032870	&  9(3,7) - 8(3,6) &	-4.57780 &	26.0 \\
91.8301065	&  9(4,5) - 8(4,4) &	-4.61854 &	29.2 \\
94.7010488	& 10(0,10) - 9(0,9) &	-4.48695 &	25.5 \\ 
\bottomrule
\end{tabular}
\label{tab:unblended}
%\addtabletext{nomenclature for the TSs refers to the numbered species in the table.}
\end{table}

%\subsection*{Supporting Information Appendix (SI)}

%Authors should submit SI as a single separate SI Appendix PDF file, combining all text, figures, tables, movie legends, and SI references. SI will be published as provided by the authors; it will not be edited or composed. Additional details can be found in the \href{https://www.pnas.org/authors/submitting-your-manuscript#manuscript-formatting-guidelines}{PNAS Author Center}. The PNAS Overleaf SI template can be found \href{https://www.overleaf.com/latex/templates/pnas-template-for-supplementary-information/wqfsfqwyjtsd}{here}. Refer to the SI Appendix in the manuscript at an appropriate point in the text. Number supporting figures and tables starting with S1, S2, etc.

%Authors who place detailed materials and methods in an SI Appendix must provide sufficient detail in the main text methods to enable a reader to follow the logic of the procedures and results and also must reference the SI methods. If a paper is fundamentally a study of a new method or technique, then the methods must be described completely in the main text.

%\subsubsection*{SI Datasets} 

%Supply .xlsx, .csv, .txt, .rtf, or .pdf files. This file type will be published in raw format and will not be edited or composed.

%\subsubsection*{SI Movies}

%Supply Audio Video Interleave (avi), Quicktime (mov), Windows Media (wmv), animated GIF (gif), or MPEG files. Movie legends should be included in the SI Appendix file. All movies should be submitted at the desired reproduction size and length. Movies should be no more than 10MB in size.

%\subsubsection*{3D Figures}

%Supply a composable U3D or PRC file so that it may be edited and composed. Authors may submit a PDF file but please note it will be published in raw format and will not be edited or composed.

\section*{Materials and Methods}

\subsection*{Astronomical Observations}

We have analysed a high-sensitivity spectral survey of the molecular cloud G+0.693-0.027 conducted with the Yebes 40m telescope (Guadalajara, Spain) and the IRAM 30m telescope (Granada, Spain).  The observations were centered at the equatorial coordinates of G+0.693: RA(J2000)=17h 47m 22s, DEC(J2000)= -28º 21’ 27”.

\subsubsection*{Yebes 40m telescope}

The observations were carried out with the Yebes 40 m telescope located in Yebes (Guadalajara, Spain), during 6 observing sessions in February 2020, as part of the project 20A008 (PI Jiménez-Serra). We used the new Nanocosmos Q-band (7 mm) HEMT receiver that enables ultra broad-band observations in two linear polarizations \cite{tercero2021}. The receiver is connected to 16 fast Fourier transform spectrometers (FFTS) with a spectral coverage of 2.5 GHz and a spectral resolution of 38 kHz. The final spectra were smoothed to a resolution of 251 kHz, corresponding to a velocity resolution of 1.7  km s$^{-1}$ at 45 GHz. We covered a total spectral range from 31.075 GHz to 50.424 GHz. The position switching mode was used, with the reference position located at (-885”,+290”) with respect to G+0.693 \cite{rivilla2020,zeng2020}. The telescope pointing and focus were checked every one or two hours through pseudo-continuum observations towards VX Sgr, a red hypergiant star near the target source. The spectra were measured in units of antenna temperature, $T_{\rm A}^*$, since the molecular emission toward G+0.693 is extended over the beams \cite{requena-torres2006}. The noise of the spectra depends on the frequency range, reaching values as low as 1.0 mK, while in some some intervals it increases up to 4.0 mK. The half power beam width ($HPBW$) of the telescope is 48” at 36 GHz.

\subsubsection*{IRAM 30m telescope}

We have carried out a spectral survey at 3 mm using the IRAM 30m telescope. The observations were performed in two observing runs during 2019: April 10-16 and August 13-19, from projects numbers 172-18 (PI Martín-Pintado), 018-19 (PI Rivilla). We used the broad-band Eight MIxer Receiver (EMIR) and the fast Fourier transform spectrometers in FTS200 mode, which provided a channel width of $\sim$200 kHz. The final spectra were smoothed to a 609 KHz, i.e. a velocity resolution of 1.8 km s$^{-1}$ at 100 GHz. The full spectral coverage is 71.770-116.720 GHz. The telescope pointing and focus were checked every 1.5 h towards bright sources.  The spectra were also measured in units of antenna temperature, $T_{\rm A}^*$. The noise of the spectra (in $T_{\rm A}^*$) is 1.3-2.8 mK in the range 71-90 GHz, 1.5-5.8 mK in the range 90-115 GHz, and $\sim$10 mK in the range 115-116 GHz. The half power beam width ($HPBW$) of the observations vary between 21.1” and 34.3”. The position switching mode was used in all observations with the off position located at (-885”,+290”) from the source position. 

\subsection*{SLIM Molecular line fitting}

The identification of the molecular lines was performed using the SLIM (Spectral Line Identification and Modeling) tool of the MADCUBA package\footnote{Madrid Data Cube Analysis on ImageJ is a software developed at the Center of Astrobiology (CAB) in Madrid: https://cab.inta-csic.es/madcuba/}. SLIM solves the radiative transfer equation, as described in detail in ref. \cite{martin2019}, and  generates the expected synthetic spectra of the molecular species under the assumption of Local Thermodynamic Equilibrium (LTE) conditions.
SLIM implements a stand-alone HyperSQL\footnote{http://hsqldb.org/} database that contains the spectral line catalogues of the Jet Propulsion Laboratory\footnote{https://spec.jpl.nasa.gov/} (JPL) \cite{pickett1998}, and the Cologne Database for Molecular Spectroscopy\footnote{https://cdms.astro.uni-koeln.de/} (CDMS) \cite{muller2001,endres2016}.

For the case of EtA, we have used the spectroscopic entry 61004 (version September 2003) of the JPL database, based on different laboratory works \cite{penn1971,kaushik1982,widicus2003}. The value of the partition function ($Q$) at the temperatures of the fit ($T_{\rm ex}\sim$ 11 K) has been interpolated from the values reported in the JPL catalog in the log$Q$-log$T$ plane, using the two adjacent temperatures: $Q$(9.375 K)=254.2935 and $Q$(18.75 K)=716.8160. 

To derive the physical parameters from the molecular emission, we have used the AUTOFIT tool of SLIM \cite{martin2019}, which performs a non-linear least squares fitting of simulated LTE spectra to the observed data. It uses the Levenberg$–$Marquardt algorithm \cite{levenberg1944,marquardt1963}, which combines the gradient descent method and the Gauss–Newton method to minimise the $\chi^2$ function.

For the analysis of EtA, we fixed the linewidth (full width at half maximum, $FWHM$) to 15 km s$^{-1}$, which reproduces well the observed spectral profiles of the EtA transitions and that is consistent with those measured for other molecules in the region \cite{zeng2018,rivilla2020,jimenez-serra2020}. We note that the upper energy levels ($E_{\rm up}$) of the transitions used in the analysis span a range between 4.8 and 29.2 K, allowing us to determine the excitation temperature ($T_{\rm ex}$) of the emission. The molecular column density ($N$), $T_{\rm ex}$ and the velocity ($v_{\rm LSR}$) were left as free parameters. The best fitting LTE model gives $N$=(1.51$\pm$0.07)$\times$10$^{13}$ cm$^{-2}$, $T_{\rm ex}$=10.7$\pm$0.7 K, and  $v_{\rm LSR}$=68.3$\pm$0.4 km s$^{-1}$.

To compute the relative molecular abundance with respect to molecular hydrogen we have used the value of the H$_2$ column density inferred from observations of C$^{18}$O, 1.35$\times$10$^{23}$ cm$^{-2}$ \cite{martin2008}. We have assumed a 20$\%$ error uncertainty in the determination of the H$_2$ column density, and propagated the error accordingly. The EtA molecular abundance falls in the range (0.9$-$1.4)$\times$10$^{-10}$.

\subsection*{Rotational diagram method}

The rotational diagram is calculated following the standard procedure \cite{goldsmith1999} implemented in MADCUBA \cite{martin2019}. For the case of optically thin emission the velocity integrated intensity over the linewidth ($FWHM$=15 km s$^{-1}$), $W$ (in K km s$^{-1}$), is converted into the column density in the upper level of the transition $N_{\rm up}$ (in cm$^{-2}$) using the expression:

\begin{equation}
N_{\rm up} = 8 \pi k \nu^2  W / (hc^3 A_{\rm ul}),
\end{equation}

where $k$ is the Boltzmann constant, $\nu$ the frequency of the transition, $h$ is the Planck’s constant, $c$ is the speed of light, and $A_{\rm ul}$ is the Einstein coefficient of spontaneous emission from the upper level $u$ to lower level $l$. Then, the level population derived for all observed transitions can be combined to determine the total molecular column density, $N$ (in cm$^{-2}$), and the excitation temperature, $T_{\rm ex}$ (in K), through the equation:

\begin{equation}
log(N_{\rm up}/g_{\rm up}) = log( N / Q(T_{\rm ex}) ) - log(e) \times E_{\rm up} / (k T_{\rm ex}),  
\end{equation}

where $g_{\rm up}$ and $E_{\rm up}$ are respectively the statistical weight and energy (in K) of the upper levels of the transitions, and $Q$ is the partition function.

Fig. \ref{fig:RD} shows the plot of log($N_{\rm up}$/$g_{\rm up}$) versus $E_{\rm up}$ for all the unblended or slightly blended transitions (see Fig. \ref{fig:eta-detections} and Table \ref{tab:unblended}). The error bars indicate the uncertainty of the velocity integrated intensity ($\Delta W$), which are derived using the expression :

\begin{equation}
\Delta W = rms \times (\Delta v / FWHM)^{0.5} \times FWHM,
\end{equation}

where $rms$ is the noise of the spectra, and $\Delta v$ is the spectral resolution of the data in velocity units. The coefficients of the straight line that fits the data points (black line in Fig. \ref{fig:RD}) provides the values for log($N$/$Q$) and log($e$)/$T_{\rm ex}$, from which MADCUBA derives $N$ and $T_{\rm ex}$, calculating $Q$($T_{\rm ex}$) as explained above. 

\subsection*{Blended transitions of ethanolamine}

We present in Fig. \ref{fig:blended} the transitions of EtA with line intensities $T_{\rm A}^*>$5 mK, as predicted by the LTE simulation described in the main text, that appear blended with emission from other molecular species already identified in the G+0.693 molecular cloud. The spectroscopic information of these transitions is shown in Table \ref{tab:blended}.

\subsection*{Tentative detection of ketenyl (HCCO) towards G+0.693-0.027}

\begin{figure*}%[tbhp]
\centering
\includegraphics[scale=0.45]{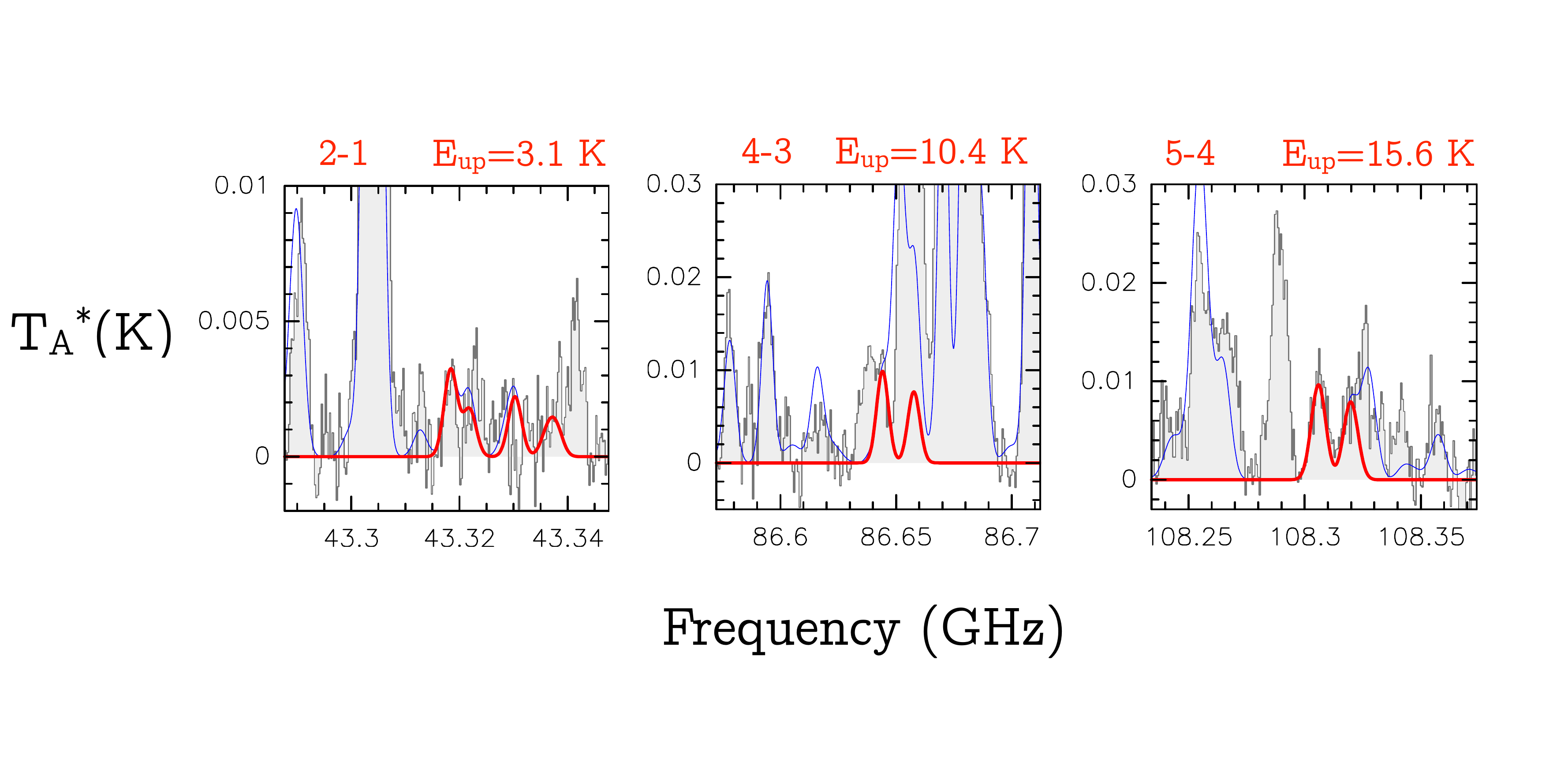}
\vspace{-1cm}
\caption{Transitions of HCCO tentatively identified in the spectra of G+0.693. The rotational quantum numbers involved in the transition are indicated in the upper left of each panel, and the energies of the upper level are indicated in the upper right. The red thick line depicts the LTE synthetic spectrum of HCCO. The blue thin line shows the predicted molecular emission from all the molecular species identified in our spectral survey, overplotted to the observed spectra (gray histograms).}
\label{fig:hcco}
\end{figure*}

We have used the CDMS entry 041506 (June 2019), based on several spectroscopic works \cite{endo1987,chantzos2019,szalay1996}. We have tentatively identified three groups of HCCO lines corresponding to the rotational transitions 2$-$1, 4$-$3, and 5$-$4. The spectra are shown in Fig. \ref{fig:hcco}, and the spectroscopic information of the transitions are listed in Table \ref{tab:hcco}. This detection should be considered tentative, since only two transitions, the 5(6,6)$-$4(5,5) and 5(6,5)$-$4(5,4) (at 108.3040553 GHz and 108.3051187 GHz, respectively) are not contaminated by emission from other species (Fig. \ref{fig:hcco}). We have produced LTE spectra using MADCUBA-SLIM and assuming $v_{\rm LSR}$=65 km s$^{-1}$ and $FWHM$=20 km s$^{-1}$. The predicted spectra reproduce well the two unblended transitions for a $T_{\rm ex}$ of 10 K, and a column density of $N\sim$0.5$\times$10$^{13}$ cm$^{-2}$ (thick red line in Fig. \ref{fig:hcco}). This column density translates into a molecular abundance of $\sim$0.4$\times$10$^{-10}$ with respect to molecular hydrogen.

%\showmatmethods{} % Display the Materials and Methods section

\acknow{
We acknowledge the constructive comments raised by the reviewers, which helped us to improve the quality of the manuscript. 
We warmly thank David Ciudad for his useful advice during the preparation of this manuscript.
We also thank Marta Ruiz Bermejo for her suggestions about several bibliographic references of ethanolamine in the context of prebiotic chemistry.
We are grateful to the IRAM 30m and Yebes 40m telescopes staff for their help during the different observing runs. IRAM is supported by INSU/CNRS (France), MPG (Germany) and IGN (Spain). The 40m radio telescope at Yebes Observatory is operated by the Spanish Geographic Institute (IGN, Ministerio de Transportes, Movilidad y Agenda Urbana). 
V.M.R. acknowledges funding support from the Comunidad de Madrid through the Atracción de Talento Investigador Modalidad 1 (contratación de doctores con experiencia) grant (COOL: Cosmic Origins Of Life; 2019-T1/TIC-15379). I.J.-S., J.M-.P. and L.C. have received partial support from the Spanish State Research Agency (AEI) through projects number PID2019-105552RB-C41, and through the Unidad de Excelencia “María de Maeztu”-Centro de Astrobiología (CSIC-INTA) project No. MDM-2017-0737. L. R. acknowledges funding support from a CSIC JAE-intro ICU studentship. P. dV and B.T. thank the support from the European Research Council through Synergy Grant ERC-2013-SyG, G.A. 610256 (NANOCOSMOS) and from the Spanish Ministerio de Ciencia e Innovación (MICIU) through project PID2019-107115GB-C21. B.T. also thanks the Spanish MICIU for funding support from grants AYA2016-75066-C2-1-P and PID2019-106235GB-I00.}

\showacknow{} % Display the acknowledgments section

\begin{table}%[tbhp]
\centering
\caption{Spectroscopic information (rest frequency, Einstein coefficients ($A_{\rm ul}$), and energy of the upper levels ($E_{\rm up}$) of the transitions of EtA that appear blended in the observed spectra of G+0.693 (see Fig. \ref{fig:blended}). All of them have line intensities $>$ 5 mK, according to the best LTE fit described in the text.}
\begin{tabular}{cccc}
Frequency (GHz) & Transition & log$A_{\rm ul}$ (s$^{-1}$)  & $E_{\rm up}$ (K) \\
\midrule
 31.7653500	&     3(1,2) - 2(1,1)	& 	-6.00114	& 	3.5    \\
 38.3746402	& 	  4(1,4) - 3(1,3)	& 	-5.71626	& 	5.1    \\
 47.8493375	&  	  5(1,5) - 4(1,4)	& 	-5.40899	& 	7.4    \\
 50.4153833	& 	  5(2,4) - 4(2,3)	& 	-5.39849	& 	9.1    \\
 73.4680803	&  	  7(2,5) - 6(2,4)	& 	-4.85633	& 	15.7    \\
 75.9126193	&  	  8(1,8) - 7(1,7)	& 	-4.78270	& 	17.0    \\
 81.4338930	& 	  8(3,6) - 7(3,5)	& 	-4.74841	& 	21.6    \\
 81.4392902	& 	  8(4,5) - 7(4,4)	& 	-4.80721	& 	24.8    \\
 81.4863522	& 	  8(4,4) - 7(4,3)	& 	-4.80646	& 	24.8    \\
 82.2698610	& 	  8(3,5) - 7(3,4)	&  	-4.73500	& 	21.7    \\
 85.1653931	& 	  9(1,9) - 8(1,8)	& 	-4.62876	& 	21.0    \\
 85.6467486	& 	  9(0,9) - 8(0,8)	& 	-4.62059	& 	21.0    \\
 91.7192567	& 	  9(4,6) - 8(4,5)	& 	-4.62007	& 	29.2    \\
 92.4964225	&  	  9(1,8) - 8(1,7)	& 	-4.52562	& 	23.0    \\
 93.0416052	& 	  9(3,6) - 8(3,5)	& 	-4.55708	& 	26.2    \\
 94.3821793	& 	10(1,10) - 9(1,9)	& 	-4.49166	& 	25.6    \\
 95.0023981	& 	  9(2,7) - 8(2,6)	& 	-4.49975	& 	24.3    \\
 99.3388437	& 	10(2,9) - 9(2,8)	& 	-4.43880	& 	28.3    \\
101.7229298	& 	10(3,8) - 9(3,7)	& 	-4.42905	& 	30.9    \\
101.8639901	& 	10(1,9) - 9(1,8)	& 	-4.39813	& 	27.9    \\
102.0183898	& 	10(4,7) - 9(4,6)	& 	-4.45929	& 	34.1    \\
102.2528632	&  	10(4,6) - 9(4,5)	& 	-4.45637	& 	34.1    \\
103.5718907	& 	11(1,11) - 10(1,10)	& 	-4.36815	& 	30.5    \\
103.7762062	& 	11(0,11) - 10(0,10)	& 	-4.36532	& 	30.5    \\
103.9759166	&  	10(3,7) - 9(3,6)	& 	-4.39964	& 	31.2    \\
105.5481523	& 	10(2,8) - 9(2,7)	& 	-4.35654	& 	29.4    \\
108.8642389	& 	11(2,10) - 10(2,9)	& 	-4.31491	& 	33.5    \\
111.0166375	& 	11(1,10) - 10(1,9)	& 	-4.28448	& 	33.2    \\
111.7757005	& 	11(3,9) - 10(3,8)	& 	-4.29721	& 	36.3    \\
112.7422238	& 	12(1,12) - 11(1,11)	& 	-4.25550	& 	35.9    \\
112.8698247	& 	12(0,12) - 11(0,11)	& 	-4.25393	& 	35.9    \\ 
\bottomrule
\end{tabular}
\label{tab:blended}
%\addtabletext{nomenclature for the TSs refers to the numbered species in the table.}
\end{table}

\begin{table}%[tbhp]
\centering
\caption{Spectroscopic information (rest frequency, Einstein coefficients ($A_{\rm ul}$), and energy of the upper levels ($E_{\rm up}$) of the rotational transitions of ketenyl (HCCO) tentatively detected towards the G+0.693 molecular cloud (shown in Fig. \ref{fig:hcco}).}
\begin{tabular}{cccc}
Frequency (GHz) & Transition & log$A_{\rm ul}$ (s$^{-1}$)  & $E_{\rm up}$ (K) \\
\midrule
 43.3176674	&   2(3,3) - 1(2,2)	&   -6.0192	& 	3.1   \\
 43.3211451	& 	2(3,2) - 1(2,1)	& 	-6.1404	& 	3.1   \\
 43.3295421	& 	2(2,2) - 1(1,1)	& 	-6.0343	& 	3.1   \\
 43.3354627	& 	2(2,1) - 1(1,0)	& 	-6.2739	& 	3.1   \\
 43.3368615	& 	2(3,2) - 1(2,2)	& 	-6.6741	& 	3.1   \\
 43.3373040	& 	2(2,1) - 1(2,1)	& 	-6.4207	& 	3.1   \\
 86.6191857	& 	4(4,3) - 3(3,3)	& 	-6.9202	& 	10.4   \\
 86.6423419	& 	4(5,5) - 3(4,4)	& 	-5.0703	& 	10.4   \\
 86.6438483	& 	4(5,4) - 3(4,3)	& 	-5.0942	& 	10.4   \\
 86.6558306	& 	4(4,4) - 3(3,3)	& 	-5.0772	& 	10.4   \\
 86.6574849	& 	4(4,3) - 3(3,2)	& 	-5.1070	& 	10.4   \\
 86.6652791	& 	4(5,4) - 3(4,4)	& 	-6.3845	& 	10.4   \\
108.2823800	& 	5(5,4) - 4(4,4)	& 	-6.7293	& 	15.6   \\
108.3040553	& 	5(6,6) - 4(5,5)	& 	-4.7698	& 	15.6   \\
108.3051187	& 	5(6,5) - 4(5,4)	& 	-4.7840	& 	15.6   \\
108.3178903	& 	5(5,5) - 4(4,4)	& 	-4.7747	& 	15.6   \\
108.3190248	& 	5(5,4) - 4(4,3)	& 	-4.7916	& 	15.6   \\
108.3280559	& 	5(6,5) - 4(5,5)	& 	-6.2997	& 	15.6    \\
\bottomrule
\end{tabular}
\label{tab:hcco}
%\addtabletext{nomenclature for the TSs refers to the numbered species in the table.}
\end{table}

% Bibliography
%\bibliography{pnas-sample}

\begin{thebibliography}{}


\bibitem[]{szostak2011}
J. W. Szostak, “An optimal degree of physical and chemical heterogeneity for the origin of life?”, Philosophical Transactions of the Royal Society B, 366, 2894-2901 (2011)

\bibitem[]{delaescosura2015}
A. de la Escosura, C. Briones, K. Ruiz-Mirazo, “The systems perspective at the crossroads between chemistry and biology”, Journal of Theoretical Biology, 381, 11-22 (2015)

\bibitem[]{deamer2005}
D. W. Deamer and J.P. Dworkin, “Chemistry and Physics of Primitive Membranes”, Topics in Current Chemistry, 259, 1-27 (2005)

\bibitem[]{sole2009}
R. V. Solé, “Evolution and self-assembly of protocells”, The International Journal of Biochemistry $\&$ Cell Biology 41, 274-84 (2009) 

\bibitem[]{hargreaves1977}
W. R. Hargreaves, S. J. Mulvihill, D. W. Deamer, “Synthesis of phospholipids and membranes in prebiotic conditions”, Nature, 266, 78-80  (1977)

\bibitem[]{oro1978}	
J. Oró, E. Sherwood, J. Eichberg, D. Epps, “Formation of phospholipids under primitive earth conditions and roles of membranes in prebiological evolution”, Deamer DW, editor. Light transducing membranes. London: Academic Press, Inc. p 1–21 (1978)

\bibitem[]{rao1987}
M. Rao, J. Eichberg, J. Oró, “Synthesis of phosphatidylethanolamine under possible primitive earth conditions”, Journal of Molecular Evolution, 25,1-6 (1987)

\bibitem[]{ruiz-mirazo2017}
K. Ruiz-Mirazo, C. Briones, A. de la Escosura,“Chemical roots of biological evolution: the origins of life as a process of development of autonomous functional systems”, Open Biology, 7 170050  (2017)

\bibitem[]{chyba1992}
C. Chyba $\&$ C. Sagan, “Endogenous production, exogenous delivery and impact shock synthesis of organic molecules: An inventory for the origins of life”, Nature, 355, 125-132 (1992)

\bibitem[]{pizzarello2010}
S. Pizzarello and E. Shock, “The Organic Composition of Carbonaceous Meteorites: The Evolutionary Story Ahead of Biochemistry”, Cold Spring Harbor perspectives in biology, 2, a002105 (2010)

\bibitem[]{bertrand2009}
M. Bertrand, S. van der Gaast, F. Vilas, F. Hörz, G. Haynes, A. Chabin, A. Brack, F. Westall, “The Fate of Amino Acids During Simulated Meteoritic Impact”, Astrobiology, 9, 943-951  (2009)

\bibitem[]{mccaffrey2014}
V. P. McCaffrey, N. E. B. Zellner, C. M. Waun, E. R. Bennett $\&$ E. K. Earl, “Reactivity and Survivability of Glycolaldehyde in Simulated Meteorite Impact Experiments”, Origins of Life and Evolution of Biospheres, 44, 29-42 (2014)

\bibitem[]{septhon2002}
M. A. Sephton, “Organic compounds in carbonaceous meteorites”, Natural Product Reports, 19, 292-311 (2002)

\bibitem[]{cooper1992}
G. W. Cooper, W. M. Onwo, J. R. Cronin, “Alkyl phosphonic acids and sulfonic acids in the Murchison meteorite”, Geochimica et Cosmochimica Acta, 56, 4109-4115 (1992)

\bibitem[]{layssac2020}
Y. Layssac, A. Gutiérrez-Quintanilla,T. Chiavassa, F. Duvernay, “Detection of glyceraldehyde and glycerol in VUV processed interstellar ice analogues containing formaldehyde: a general formation route for sugars and polyols”, Monthly Notices of the Royal Astronomical Society, 496, 5292-5307 (2020)

\bibitem[]{zhu2020}
C. Zhu, A. M. Turner, M. J. Abplanalp, R. I. Kaiser, B. Webb, G. Siuzdak, R. C. Fortenberry,  “An Interstellar Synthesis of Glycerol Phosphates”, The Astrophysical Journal Letters, 899, L3 (2020)

\bibitem[]{zhang2017}
X. Zhang, G. Tian, J. Gao, M. Han, R. Su, Y. Wang, S. Feng, “Prebiotic Synthesis of Glycine from Ethanolamine in Simulated Archean Alkaline Hydrothermal Vents”, Origins of Life and Evolution of Biospheres, 47, 413-425 (2017) 

\bibitem[]{sojo2016}
V. Sojo, B. Herschy, A. Whicher, E. Camprubí, N. Lane, “The Origin of Life in Alkaline Hydrothermal Vents”, Astrobiology 16, 181-197 (2016)

\bibitem[]{glavin2010}
D. P. Glavin, A. D. Aubrey, M.l P. Callahan, J P. Dworkin, J. E. Elsila, E. T. Parker, J. L. Bada, P. Jennisken, M. H. Shaddad, “Extraterrestrial amino acids in the Almahata Sitta meteorite”, Meteoritics $\&$ Planetary Science, 45, 1695-1709 (2010)

\bibitem[]{wirstrom2007}
E.S. Wirström, P. Bergman, A. Hjalmarson, A. Nummelin, “A search for pre-biotic molecules in hot cores”, Astronomy $\&$ Astrophysics, 473, 177-180 (2007)

\bibitem[]{requena-torres2008}
M.A. Requena-Torres, J. Martín-Pintado, S. Martín, M. R. Morris, “The Galactic Center: The Largest Oxygen-bearing Organic Molecule Repository”, The Astrophysical Journal, 672, 352-360 (2008)

\bibitem[]{zeng2018}
S. Zeng, I. Jiménez-Serra, V.M Rivilla, S. Martín, J. Martín-Pintado, M.A. Requena-Torres, J. Armijos-Abendaño, D. Riquelme, R. Aladro, “Complex organic molecules in the Galactic Centre: the N-bearing family”,  Monthly Notices of the Royal Astronomical Society, 478, 2962-2975 (2018)

\bibitem[]{rivilla2019}
V. M. Rivilla, J. Martín-Pintado, I. Jiménez-Serra, S. Zeng, S. Martín, J. Armijos-Abendaño, M. A. Requena-Torres, R. Aladro, D Riquelme, “Abundant Z-cyanomethanimine in the interstellar medium: paving the way to the synthesis of adenine”, Monthly Notices of the Royal Astronomical Society: Letters, 483, 114-119 (2019)

\bibitem[]{rivilla2020}
V. M. Rivilla, J. Martín-Pintado, I. Jiménez-Serra, S. Martín, L. F. Rodríguez-Almeida, M. A. Requena-Torres, F. Rico-Villas, S. Zeng, C. Briones, “Prebiotic Precursors of the Primordial RNA World in Space: Detection of NH2OH”, The Astrophysical Journal Letters, 899, L28 (2020)

\bibitem[]{jimenez-serra2020}
I. Jiménez-Serra, J. Martín-Pintado, V.M. Rivilla, L. F. Rodríguez-Almeida, E.R. Alonso, S. Zeng, E.J. Cocinero, Emilio, S. Martín, M.A. Requena-Torres, R. Martín-Domenech, L. Testi, “Toward the RNA-World in the Interstellar Medium—Detection of Urea and Search of 2-Amino-oxazole and Simple Sugars”, Astrobiology, 20, 1048-1066 (2020)

\bibitem[]{martin2008}
S. Martín, M. A. Requena-Torres, J. Martín-Pintado, R. Mauersberger, “Tracing Shocks and Photodissociation in the Galactic Center Region”, The Astrophysical Journal, 678, 245-254 (2008)

\bibitem[]{zeng2020}
S. Zeng, Q. Zhang, I. Jiménez-Serra, B. Tercero, X. Lu, J. Martín-Pintado, P. de Vicente, V.M. Rivilla, S. Li, “Cloud-cloud collision as drivers of the chemical complexity in Galactic Centre molecular clouds”, Monthly Notices of the Royal Astronomical Society, 497, 4896-4909 (2020)

\bibitem[]{martin2019}
S. Martín, J. Martín-Pintado, C. Blanco-Sánchez, V.M. Rivilla, A. Rodríguez-Franco, F. Rico-Villas, F.,  “Spectral Line Identification and Modelling (SLIM) in the MAdrid Data CUBe Analysis (MADCUBA) package. Interactive software for data cube analysis” , Astronomy $\&$ Astrophysics, 631, A159, 17pp. (2019) 

\bibitem[]{mcguire2018}
B. A. McGuire “2018 Census of Interstellar, Circumstellar, Extragalactic, Protoplanetary Disk, and Exoplanetary Molecules”, The Astrophysical Journal Supplement Series, 239, Issue 2, article id. 17, 48pp (2018)

\bibitem[]{belloche2008}
A. Belloche, K. M. Menten, C. Comito, H. S. P. Müller, P. Schilke, J. Ott, S. Thorwirth, C. Hieret,  “Detection of amino acetonitrile in Sgr B2(N)”, Astronomy and Astrophysics, 482, 179-196 (2008)

\bibitem[]{zeng2019}
S. Zeng, D. Quénard, I. Jiménez-Serra, J. Martín-Pintado, V.M. Rivilla, L. Testi, R. Martín-Doménech, “First detection of the prebiotic molecule glycolonitrile (HOCH2CN) in the interstellar medium”, Monthly Notices of the Royal Astronomical Society: Letters, 484, L43-L48 (2019)	

\bibitem[]{bernstein2002}
M. P. Bernstein, J. P. Dworkin, S. A. Sandford, G. W. Cooper, L. J. Allamandola, “Racemic amino acids from the ultraviolet photolysis of interstellar ice analogues”, Nature, 416, 401-403 (2002)


\bibitem[]{ehrenfreund2001}
P. Ehrenfreund $\&$ S. B. Charnley, “From astrochemistry to astrobiology”, Proceedings of the First European Workshop, 21 - 23 May 2001, Frascati, Italy. Eds.: P. Ehrenfreund, O. Angerer $\&$ B. Battrick. ESA SP-496, Noordwijk: ESA Publications Division,, p. 35 - 42 (2001)

\bibitem[]{charnley2001}
S. B. Charnley, "Interstellar organic chemistry", 2001: The Bridge between the Big Bang and Biology, F. Giovannelli (ed.), CNR President Bureau, Spatial Volume, 139-149, Roma (2001)

\bibitem[]{agundez2015}
M. Agúndez, J. Cernicharo, M. Guélin, “Discovery of interstellar ketenyl (HCCO), a surprisingly abundant radical”, Astronomy $\&$ Astrophysics, 577, L5, pp. 6 (2015)

\bibitem[]{wakelam2015}
V. Wakelam, V., J. -C  Loison, K. M. Hickson, M. Ruaud, "A proposed chemical scheme for HCCO formation in cold dense clouds", Monthly Notices of the Royal Astronomical Society: Letters, 453, L48-L52 (2015)

\bibitem[]{bizzocchi2020}
L. Bizzocchi, D. Prudenzano, V. M. Rivilla, A. Pietropolli-Charmet, B. M. Giuliano, P. Caselli, J. Martín-Pintado, I. Jiménez-Serra, S. Martín, M.A. Requena-Torres, F. Rico-Villas, S. Zeng, J.-C. Guillemin, “Propargylimine in the laboratory and in space: millimetre-wave spectroscopy and its first detection in the ISM”, Astronomy $\&$ Astrophysics, 640, A98, pp. 14 (2020)

\bibitem[]{kameneva2017}
S. V. Kameneva, D. A. Tyurin, V. I. Feldman, "Characterization of the HCN-CO complex and its radiation-induced transformation to HNC-CO in cold media: an experimental and theoretical investigation", Phys. Chem. Chem. Phys., 19, 24348-24356 (2017)

\bibitem[]{krasnokutski2020}
S. A. Krasnokutski, "Did life originate from low-temperature areas of the Universe?", eprint arXiv:2010.10905 (2020)

\bibitem[]{tanaka2011}
K. Tanaka, T. Oka, S. Matsumura, M. Nagai, K. Kamegai, "High atomic carbon abundance in molecular clouds in the Galactic Cneter region", The Astrophysical Journal Letters, 743, L39

\bibitem[]{suzuki2018}
T. Suzuki, L. Majumdar, M. Ohishi, M. Saito, T. Hirota, V. Wakelam, "An Expanded Gas-grain Model for Interstellar Glycine",  
The Astrophysical Journal, 863, article id. 51 (2018).

\bibitem[]{ioppolo2020}
S. Ioppolo, G. Fedoseev, K.-J. Chuang, H.M. Cuppen, A.R. Clements, M. Jin, R.T. Garrod, D. Qasim, V. Kofman, E.F. van Dishoeck and H. Linnartz, “A non-energetic mechanism for glycine formation in the interstellar medium”, Nature Astronomy (2020)

\bibitem[]{fedoseev2015}
G. Fedoseev, H. M. Cuppen, H. M., S. Ioppolo, T. Lamberts, H. Linnartz. “Experimental evidence for glycolaldehyde and ethylene glycol formation by surface hydrogenation of CO molecules under dense molecular cloud conditions”, Monthly Notices of the Royal Astronomical Society, 448, 1288-1297 (2015)


\bibitem[]{garrod2006}
R. T. Garrod $\&$ E. Herbst, "Formation of methyl formate and other organic species in the warm-up phase of hot molecular cores", Astronomy $\&$ Astrophysics,457, 927-936 (2006)

\bibitem[]{garrod2008}
R. T. Garrod, S. L. Widicus Weaver, E. Herbst, "Complex Chemistry in Star-forming Regions: An Expanded Gas-Grain Warm-up Chemical Model", The Astrophysical Journal, 682, 283-302 (2008)

\bibitem[]{garrod2013}
R. T Garrod, "A Three-phase Chemical Model of Hot Cores: The Formation of Glycine", The Astrophysical Journal, 765, article id. 60 (2013).

\bibitem[]{ruaud2015}
M. Ruaud, J. -C. Loison, K. M. Hickson, P. Gratier, F. Hersant, V. Wakelam, "Modelling complex organic molecules in dense regions: Eley-Rideal and complex induced reaction", Monthly Notices of the Royal Astronomical Society, 447, 4004-4017 (2015)


\bibitem[]{redondo2020}
P. Redondo, M. Sanz-Novo, A. Largo, C. Barrientos, "Amino acetaldehyde conformers: structure and spectroscopic properties", Monthly Notices of the Royal Astronomical Society, 492, 1827-1833 (2020)


\bibitem[]{whittet1991}
D.C.B. Whittet and W.W. Duley, “Carbon monoxide frosts in the interstellar medium“, The Astronomy and Astrophysics Review, 2, 167-189 (1991)

\bibitem[]{goodrich2015}
C. A. Goodrich, W. K. Hartmann, D. P. O’Brien, S. J. Weidenschilling, L. Wilson, P. Michel, M. Jutzi, “Origin and history of ureilitic material in the solar system: The view from asteroid 2008 TC3 and the Almahata Sitta meteorite”, Meteoritics $\&$ Planetary Science 50, 4, 782–809 (2015)

\bibitem[]{piani2020}
L. Piani, Y. Marrocchi, T. Rigaudier, L. G. Vacher, D. Thomassin, B. Marty, “Earth’s water may have been inherited from material similar to enstatite chondrite meteorites”, Science , 369, 1110-1113 (2020)

\bibitem[]{monnard2002}
P. A. Monnard and D. W. Deamer, “Membrane self‐assembly processes: Steps toward the first cellular life”  (2002), The Anatomical Record, 268, 196-207 (2002)

\bibitem[]{budin2011}	
I. Budin and J. W. Szostak, “Physical effects underlying the transition from primitive to modern cell membranes” , Proceedings of the National Academy of Sciences of the United States of America, 108, 5249-5254 (2011)

\bibitem[]{tercero2021}
F. Tercero, J. A. López-Pérez, J. D.. Gallego, F. Beltrán, O. García, M. Patino-Esteban, I. López-Fernández, G. Gómez-Molina, M. Diez, P. García-Carreño, I. Malo, R. Amils, J.M. Serna, C. Albo, J. M. Hernández, B. Vaquero, J. González-García, L. Barbas, J. A. López-Fernández, V. Bujarrabal, M. Gómez-Garrido, J. R. Pardo, M. Santander-García, B. Tercero, J. Cernicharo, P. de Vicente, “Yebes 40 m radio telescope and the broad band NANOCOSMOS receivers at 7 mm and 3 mm for line surveys”, Astrpnomy $\&$ Astrophysics, 645, A37 (2021)

\bibitem[]{requena-torres2006}
M. A. Requena-Torres, J. Martín-Pintado, A. Rodríguez-Franco, S. Martín, N. J. Rodríguez-Fernández, P. de Vicente, “Organic molecules in the Galactic center. Hot core chemistry without hot cores”, Astronomy $\&$ Astrophysics, 455, 971-985 (2006)

\bibitem[]{pickett1998}
H. M. Pickett, R. L. Poynter, E. A. Cohen, M. L. Delitsky, J. C. Pearson, H. S. P. Müller, “Submillimeter, millimeter and microwave spectral line catalog”, Journal of Quantitative Spectroscopy and Radiative Transfer, 60, 883-890 (1998)

\bibitem[]{muller2001}
H. S. P Müller, S. Thorwirth, D. A. Roth, G. Winnewisser, “The Cologne Database for Molecular Spectroscopy, CDMS”, Astronomy and Astrophysics, 370, L49-L52 (2001) 

\bibitem[]{endres2016}
C. P. Endres, S. Schlemmer, P. Schilke, J. Stutzki, H. S. P. Müller, “The Cologne Database for Molecular Spectroscopy, CDMS, in the Virtual Atomic and Molecular Data Centre, VAMDC”, Journal of Molecular Spectroscopy, 327, 95-104  (2016)

\bibitem[]{penn1971}
R. E. Penn, R. F. J. Curl, “Microwave Spectrum of 2‐Aminoethanol: Structural Effects of the Hydrogen Bond”, The Journal of Chemical Physics, 53, 651-658 (1971)

\bibitem[]{kaushik1982}
V. K. Kaushik, R.C. Woods, “Centrifugal Distortion Effects in the Rotational Spectrum of 2-Aminoethanol”, Zeitschrift fur Physikalische Chemie Neue Folge, 132, 117-120 (1982)

\bibitem[]{widicus2003}
S. L. Widicus Weaver, B. J. Drouin, K. A. Dyl, and G. A. Blake, “Millimeter wavelength measurements of the rotational spectrum of 2-aminoethanol”, Journal of  Molecular Spectroscopy, 217, 278-281 (2003)

\bibitem[]{levenberg1944}
K. Levenberg, “A method for the solution of certain non-linear problems in least squares”, Quarterly of Applied Mathematics, 2, 164-168 (1944)

\bibitem[]{marquardt1963}
D. Marquardt, "An Algorithm for Least-Squares Estimation of Nonlinear Parameters". Journal of the Society for Industrial and Applied Mathematics, 11, 431-441 (1963)

\bibitem[]{goldsmith1999}
P. F. Goldsmith $\&$ W. D. Langer “Population Diagram Analysis of Molecular Line Emission”, The Astrophysical Journal, 517, 209-225 (1999)

\bibitem[]{endo1987}
Y. Endo and E. Hirota, “The submillimeter‐wave spectrum of the HCCO radical”, The Journal of Chemical Physics, 86, 4319-4326 (1987)

\bibitem[]{chantzos2019}
J. Chantzos, S. Spezzano, C. Endres, L. Bizzocchi, V. Lattanzi, J. Laas, A. Vasyunin, P. Caselli, “Rotational spectroscopy of the HCCO and DCCO radicals in the millimeter and submillimeter range”, Astronomy $\&$ Astrophysics, 621, A111, pp. 7 (2019) 

\bibitem[]{szalay1996}
P. G. Szalay, G. Fogarasi, L. Nemes, “Quantum chemical coupled cluster study of the structure and spectra of the ground and first excited states of the ketenyl radical”, Chemical Physics Letters, 263, 91-99 (1996)




\end{thebibliography}

\end{document}